\documentclass[prb,twocolumn,amsmath,amsfonts,amssymb]{revtex4}
\usepackage{graphicx}
\usepackage {simplewick}
\usepackage{color}
\usepackage{here}
\usepackage{bm}

\allowdisplaybreaks[4]
\newcommand{\bra}{\langle}
\newcommand{\ket}{\rangle}

\newcommand{\bre}{\nonumber\\}

\definecolor{purple}{rgb}{0.8,0.0,0.8}

\def\be{\begin{equation}}
\def\ee{\end{equation}}
\def\bea{\begin{eqnarray}}
\def\eea{\end{eqnarray}}

\begin{document}

\title{Configuration interaction combined with spin-projection for strongly correlated molecular electronic structures}
\author{Takashi Tsuchimochi}
\email{tsuchimochi@gmail.com}
\author{Seiichiro Ten-no}
\email{tenno@garnet.kobe-u.ac.jp}
\affiliation{Graduate School of System Informatics, Kobe University, Kobe 657-8501 Japan}
\begin{abstract}
We introduce single and double particle-hole excitations in the recently revived spin-projected Hartree-Fock. Our motivation is to treat static correlation with spin-projection and recover the residual correlation, mostly dynamic in nature, with simple configuration interaction (CI). To this end, we present the Wick theorem for nonorhtogonal determinants, which enables an efficient implementation in conjunction with the direct CI scheme. The proposed approach, termed ECISD, achieves a balanced treatment between dynamic and static correlations. To approximately account for the quadruple excitations, we also modify the well-known Davidson correction. We report our approaches yield surprisingly accurate potential curves for HF, H$_2$O, N$_2$, and a hydrogen lattice, compared to traditional single reference wave function methods at the same computational scaling as regular CI.  
\end{abstract} 
\maketitle

Despite the long elaboration of theoretical developments for electron correlation, an accurate black-box description of dynamic and static (strong) correlations yet remains unreached. The difficulty comes from the fact that most multi-reference (MR) methods require to define an active space, where electrons are assumed to be strongly correlated.\cite{Roos80, Siegbahn80,Siegbahn81} If such a space is not appropriately set up, one does not obtain a qualitatively correct zeroth order wave function and perturbation theory can suffer from intruder states.\cite{Andersson94} Introducing a larger active space to avoid this issue increases the computational cost exponentially, and thus its application has been limited to small systems. 

Recently, many authors have vigorously tackled the strong correlation problem from different perspectives.\cite{White92, Chan02, Chan11, Gidofalvi08, Tsuchimochi09,Scuseria09, Scuseria11,Jimenez12,Limacher13,Stein14,Henderson14,Phillips14} 
Among them, projected Hartree-Fock (PHF) stands on symmetry-breaking and restoration in the mean-field picture, and delivers static correlations efficiently.\cite{Lowdin55,Jimenez12} Scuseria {\it et al.} have extended PHF to describe the residual dynamic correlations by sequentially updating and adding PHF wave functions based on approximate excited PHF states.\cite{Rodriguez13,Jimenez13B} Hence, the approach employs different orbitals for different configurations, and is commonly regarded as nonorthogonal configuration interaction (NOCI). The spirit behind NOCI is that the energy convergence with respect to the number of configurations is much faster than the orthogonal one. Unfortunately, the number of configurations required to achieve the chemical accuracy  in NOCI increases exponentially with the system size,\cite{LeBlanc15} hampering its practical applications in part because it is hard to know when to terminate the CI expansion, which is crucial for a balanced treatment between different systems. 

A more natural choice for CI basis is the ``particle-hole'' picture of excited determinants. While conceptually simple, it was considered that this canonical CI scheme would involve a very high cost due to the lack of simple Slater-Condon rule for the projected (nonorthogonal) Hamiltonian elements, whose evaluations thus were determined to be the most computationally demanding step.\cite{Gao09, Jimenez13B} In this manuscript, however, we show that one can formulate the particle-hole CI in a manner quite similar to the regular direct CI scheme\cite{Roos72} while retaining the same computational scaling.

Our wave function {\it ansatz} is therefore CI with singles and doubles (CISD) built from the broken-symmetry deformed state $|\Phi\ket$, which is explicitly projected to the correct symmetry-space with projection operator $\hat P$,
\begin{align}
|\Psi\ket = \hat P \left(c_0|\Phi\ket + \sum_{ia} c_i^a |\Phi_i^a\ket + \sum_{i<j, a<b}c_{ij}^{ab} |\Phi_{ij}^{ab}\ket \right),\label{eq:ECISD}
\end{align}
where $|\Phi_i^a\ket$ and $|\Phi_{ij}^{ab}\ket$ are singly and doubly excited determinants. Throughout this Communication, we will stick with the conventional orbital indices, where $i,j,k,l$ represent occupied, $a,b,c,d$ virtual, and $p,q,r,s$ general orbitals. Also, we only consider the $\hat S^2$-symmetry in our projection because it is  widely recognized as the most important symmetry for static correlation. $\hat P$ is thus an exact spin-projector and removes {\it all} the spin-contaminants associated with broken symmetry determinants in the CI expansion, so that $|\Psi\ket$ is an eigenfunction of $\hat S^2$. 
Therefore, we will henceforth call this scheme spin-extended CISD (ECISD), in the philosophy of L\"owdin's spin-extended HF,\cite{Lowdin55} which is also referred to as spin-projected unrestricted HF (SUHF) in the recent literature.\cite{Jimenez12} Although ECISD employs orthogonal broken-symmetry determinants, it  can be also considered  a NOCI method with well-defined coordinates, $\{\hat P |\Phi\ket, \hat P |\Phi_i^a\ket, \hat P|\Phi_{ij}^{ab}\ket\}$, and the basis is generically overcomplete. 

The residual correlation of ECISD, $\Delta E_{\rm SD} = E_{\rm ECISD} - E_{\rm SUHF}$, is given by
\begin{align}
\Delta E_{\rm SD} = \frac{\bra \Psi |\bar H |\Psi \ket}{\bra \Psi | \Psi\ket}
\end{align}
where $\bar H = \hat H - E_{\rm SUHF}$. As in regular CISD, one variationally minimizes $\Delta E_{\rm SD}$, resulting in the following equations for configuration coefficients ${\bf c}$:
\begin{align}
&\bra \nu | \bar H |\Psi \ket = \Delta E_{\rm SD} \bra \nu | \Psi \ket \;\;\;\;\;\; \; \; \forall \;\;\; \nu \in \{\Phi, \Phi_i^a, \Phi_{ij}^{ab}\} \label{eq:ECISDeq}
\end{align}
where we have used $\hat P = \hat P^2 = \hat P^\dag$ and $[\hat H, \hat P] = 0$. This set of equations can be solved as a generalized eigenvalue problem,
\begin{align}
\bar {\bf H}{\bf c} = \Delta E_{\rm SD}{\bf S}{\bf c}\label{eq:HCSCE}
\end{align}
where $\bar {\bf H}$ and ${\bf S}$ are the Hamiltonian and overlap matrix elements in the basis of projected determinants. 

As is widely known, it is not practical to employ L\"owdin's spin-projection operator even for a single broken-symmetry determinant, because the operator in this form is characterized as an intractable many-body manifold. Obviously, each excited determinant is further spin-contaminated compared to $|\Phi\ket$, and therefore the use of this traditional operator would make ECISD completely infeasible. Instead, we use the integration scheme with one-body spin-rotation operator $\hat R(\Omega)$,\cite{RingSchuck, Scuseria11,Jimenez12} 
\begin{align}
\hat P = \int_\Omega d\Omega  \;\;w(\Omega) \hat R(\Omega) \simeq \sum_g^{N_{\rm grid}} w_g \hat R_g\label{eq:P}
\end{align} 
where the numerical integration usually requires 7$\sim$10 grid points. For more details, see Ref.[\onlinecite{Jimenez12}]. 

Given Eqs.(\ref{eq:ECISDeq}) and (\ref{eq:P}), our task is to formulate the matrix elements where ket is unitarily rotated with $\hat R_g$. Hence, it is convenient to introduce unitarily transformed fermions,
\begin{align}
b^\dag_p = \hat R_g a_p^\dag \hat R_g^{-1} = \sum_q R_{qp} a_q^\dag,
\end{align}
with $R_{pq} = \bra \phi_p |\hat R_g | \phi_q\ket$.
Note $\hat R_g | \Phi\ket \equiv \prod_{i} b^\dag_i |-\ket$ is a single Slater determinant nonorthogonal to $\Phi$ with the physical vacuum $|-\ket$, and fermions $b_p$ depend on $g$.
For our present purpose, we need to evaluate the transition density matrix element of an arbitrary string in the form, 
\begin{align}
&\frac{\bra \Phi| a^\dag_{i_1} \cdots a^\dag_{i_M} a_{a_M} \cdots a_{a_1}  \hat R_g  a_{b_1}^\dag \cdots a_{b_N}^\dag a_{j_N}\cdots a_{j_1} |\Phi\ket}{\bra \Phi|\hat R_g |\Phi\ket} \bre
&= \bra a^\dag_{i_1} \cdots  a^\dag_{i_M}  a_{a_M} \cdots a_{a_1} b^\dag_{b_1} \cdots b^\dag_{b_N} b_{j_N} \cdots b_{j_1}\ket_g, \label{eq:kPDM}
\end{align} 
where we introduced the short hand notation,
\be
\bra \cdots \ket_g \equiv \frac{\bra \Phi| \cdots \hat R_g |\Phi\ket}{\bra \Phi|\hat R_g |\Phi\ket}.
\ee
The seminal works of L\"owdin\cite{Lowdin55} showed the way to calculate the transition density matrix of an arbitrary operator in terms of the one-body transition density matrix as a fundamental invarianant.
Nevertheless, the formulation based on the first quantization is not fully convenient to construct many-body theory including the spin projection operators.
Inspired by L\"owdin's invariant, we develop the nonorthogonal Wick theorem to efficiently evaluate the transition density matrix elements in what follows.

Similarly to the extended Wick theorem for the MR case,\cite{Kutzelnigg97} a normal ordered operator for two nonorthogonal determinants can be expressed by subtracting the expectation values order-by-order in the operator rank.
The fact that the expectation value for the nonorthogonal Slater-determinants is factrizable into one-body transition density matrices may allow us to express the normal ordered operator in exactly the same manner as the single-reference one\cite{Lindgren} in an orthogonal basis,
\be
ABC\cdots = \{ABC\cdots \} + \sum_{\rm all \; contractions} \{\contraction[0.7ex]{}{A}{BC\;}{}\contraction[1.4ex]{A}{B}{C\;\;\;}{}ABC\cdots\}
\ee
where $\{\cdots\}$ denotes a normal ordered operator. 
 Especially for Eq.(\ref{eq:kPDM}), we only need to consider the following contractions which give nonzero expectation values,
\begin{align}
{\bm{\mathcal W}}_g := \begin{pmatrix}
\contraction[0.7ex]{}{a_o^\dag}{}{b} a_o^\dag b_o & \contraction[0.7ex]{}{a_o^\dag}{}{a} a_o^\dag a_v \\
\contraction[0.7ex]{}{b_v}{}{b} b_v^\dag b_o & \contraction[0.7ex]{}{a_v}{}{b} a_v b_v^\dag
\end{pmatrix}
=
\begin{pmatrix}
\bra a_o^\dag b_o\ket_g & \bra a^\dag_o a_v\ket_g \\
\bra b_v^\dag b_o\ket_g & \bra a_v b_v^\dag\ket_g
\end{pmatrix}, \label{eq:Wg}
\end{align}
in addition to the usual contractions for the fermions, $\contraction[0.7ex]{}{a_i^\dag}{}{a} a_i^\dag a_j  = \contraction[0.7ex]{}{b_i^\dag}{}{b} b_i^\dag b_j = \delta_{ij}$.
Here $o$ and $v$ indicate the occupied and virtual spaces of the molecular orbitals of $|\Phi\ket$. ${\bm{\mathcal W}}_g$ becomes the identity matrix in the case of regular single reference methods, thus being unique to nonorthogonal methods.
Then, the projected matrix elements can be solely expressed by ${\bm{\mathcal W}}_g$ (along with $\delta$ functions for $\bra a^\dag_o a_o\ket_g$ and $\bra b_o^\dag b_o \ket_g$) using the nonorthogonal Wick theorem. 
This fact enables an efficient construction of $\bar {\bf H}{\bf c}$ and ${\bf Sc}$ with computational scalings of ${\cal O}(o^2v^4)$ and ${\cal O}(o^2v^3)$ for each $g$, respectively, allowing practical calculations with direct CI.\cite{Roos72} The explicit working equations are given in Supporting Information,\cite{SI} and the detailed formulation will be given in the future work.

One of the major disadvantages of PHF methods, including SUHF, is that they are not size-consistent\cite{Scuseria11, Jimenez12, Henderson13, Tsuchimochi15} except for some special cases.\cite{Tsuchimochi11, Tsuchimochi15C} Therefore, one of the most desirable features we expect for the residual correlation effect in post-PHF methods\cite{Rodriguez13,Jimenez13B,Tsuchimochi14} is size-consistency and, if possible, the capability of removing, or at least mitigating the size-consistency error resulting from PHF.  However, since ECISD truncates the full-CI (FCI) expansion at doubles and neglects simultaneous double excitations, its correlation energy obviously does not have the proper scaling with respect to the system size.\cite{Duch94} To correct this, we consider, in analogy with regular CISD and MRCI, an {\it a posteriori} treatment of approximate quadruple excitations using the Davidson correction scheme.\cite{Davidson74, Langhoff74} Generally, a variety of the Davidson corrections require the CI coefficient $c_0$ of the reference wave function, e.g., the simplest one being $\Delta E_{\rm Q} = (1-c_0^2) \Delta E_{\rm SD}.$\cite{Duch94}
In ECISD, on the other hand, this equation is not directly applicable because our reference wave function $\hat P|\Phi\ket$ permeates the singles and doubles spaces through $\hat P$, as manifested by the fact $\bra \Phi_i^a | \hat P |\Phi\ket \ne 0$. 

The role of $c_0^2$ in the Davidson corrections for (MR)CISD is that it carries the information about the {\it weight} that the reference state occupies in the space spanned by $|\Psi\ket$. Hence, one can introduce operator $\hat {\cal O} = \hat P| \Phi\ket \bra \Phi | \hat P |\Phi \ket^{-1}\bra \Phi | \hat P^\dag$, which performs projection onto the SUHF subspace. A ``$c_0^2$--like'' quantity can then be simply defined as the expectation value of $\hat {\cal O}$, and the Davidson correction becomes
\begin{align}
\Delta E_{\rm Q} 
&= \left( 1- \frac{|\bra \Phi | \Psi\ket|^2}{\bra \Phi|\hat P |\Phi\ket \bra \Psi|\Psi\ket} \right )\Delta E_{\rm SD}.\label{eq:DC2}
\end{align}
While there are several variants of {\it a posteriori} size-consistent corrections,\cite{Duch94} we found that most schemes gave similar results for the molecules we have tested below, and hence we will only report the results with Eq.(\ref{eq:DC2}). 

Before proceeding to the numerical results, we describe the computational details. 
 In our implementation of ECISD, the diagonalization of Eq.(\ref{eq:HCSCE}) is carried out with direct CI employing the iterative technique due to Davidson\cite{Davidson75} extended to a generalized eigenvalue problem. 
ECISD with the simple diagonal preconditioning typically takes three-four times the number of cycles  compared to  regular CISD.\cite{Tsuchimochi15} We will address ways to ameliorate this slow convergence in a forthcoming paper.  We use the converged SUHF orbitals within a variation-after-projection scheme, which are then fixed during the ECISD calculations. 
In the calcualtions of N$_2$, $1s$ orbitals are frozen to enable the direct comparison against the exact FCI results. The active spaces used in CASPT2\cite{Andersson94}  and internally-contracted MRCI\cite{Werner88,Knowles88}  are $(2e,2o)$ for HF, $(6e,5o)$ for H$_2$O, and $(6e,6o)$ for N$_2$. We have used a 6-31G basis for all the calculations except for the H$_{12}$ plane where a STO-3G basis was employed.  Although regular CISD is obsolete, we also included it in our discussion to show the improvement ECISD has to offer. 

\begin{figure}[t!]
\includegraphics[width=80mm]{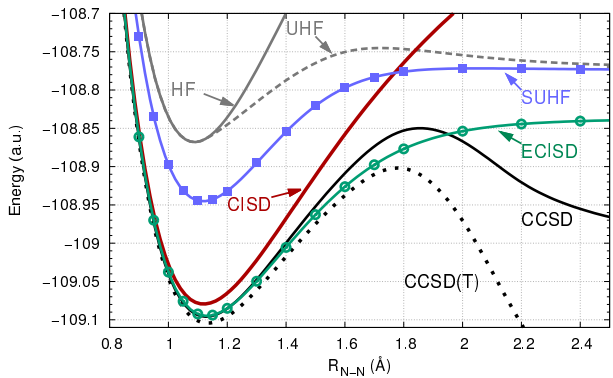}
\caption{Potential energy curves of N$_2$ with a 6-31G basis.}\label{fig:N2_E}
\end{figure}

We first demonstrate the performance of ECISD using the dissociation of the N$_2$ molecule, which still largely remains challenging for major single reference methods such as CCSD. This system requires a considerable amount of dynamic correlation in the vicinity of equilibrium while the correct treatment of static correlation is indispensable as the internuclear distance increases. This is clearly seen in Figure \ref{fig:N2_E}; all the tested single-reference methods bring dynamic correlation and drastically improve the description over HF around equilibrium, but become inaccurate when the molecule is pulled apart. This deficiency is mostly attributed to the inadequacy of the HF reference.
Energetically correct pictures for static correlation may be alternatively obtained by breaking the spin-symmetry of a wave function at the price that the resultant unrestricted wave function is not an eigenfunction of the spin operator. Therefore, while unrestricted HF (UHF) depicts the correct shape of the dissociation curve, many physical properties such as excitation energies are vastly incorrect.\cite{Tsuchimochi15} On the other hand, SUHF restores the spin-symmetry in the reference broken-symmetry HF determinant $|\Phi\ket$ and describes the dissociation limit correctly; however, it significantly underestimates the correlation energy due to the lack of major dynamic correlation effects. ECISD delivers the residual correlation that is missing in SUHF, while retaining the correct description of static correlation. Noteworthy is that, not only at the dissociation limit where static correlation is abundant, but also at the equilibrium distance, ECISD outperforms CISD, and is as accurate as CCSD: we deem this is simply because even around this region SUHF provides a much better reference than does HF as a starting point.

\begin{figure*}[t!]
\includegraphics[width=170mm]{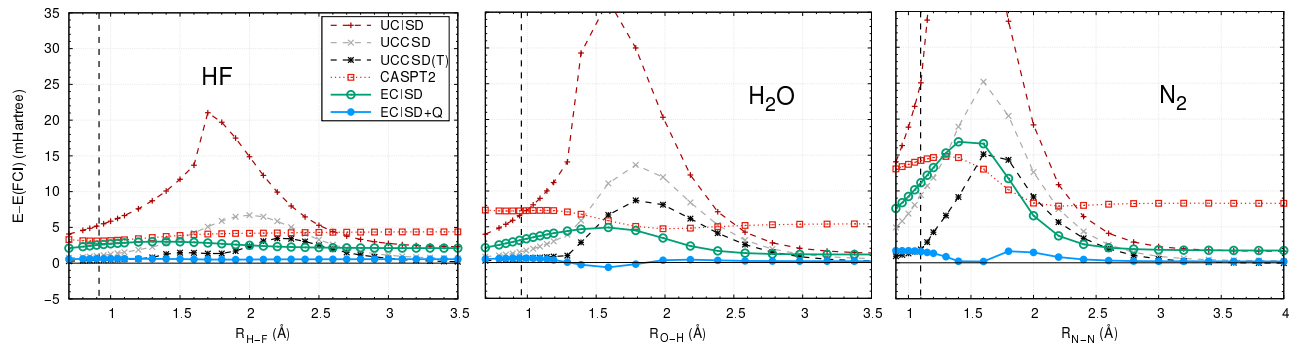}
\caption{Energy differences $E-E({\rm FCI})$ for the bond-breaking curves for HF, H$_2$O (symmetric dissociation), and N$_2$.  Vertical dashed lines indicate the equilibrium distances.}\label{fig:curves}
\end{figure*} 

Figure \ref{fig:curves} illustrates the errors in mHartree from the FCI energy for the potential curves of HF, the symmetrically stretched H$_2$O at $\angle$HOH$=109.57^\circ$, and N$_2$. These three examples showcase the hierarchy of entanglements; two, four, and six electrons are strongly correlated, respectively, and an accurate description becomes harder to achieve in this order. For comparison, the results of unrestricted methods and CASPT2 are also plotted. Vertical dashed lines are added at experimental R$_{\rm e}$ as an eye guide, and the non-parallelity errors (NPE), the absolute difference between the maximum and minimum deviations from the FCI energy, are summarized in Table \ref{tb:NPE}. Again, ECISD achieves a significant improvement over UCISD at the same computational scaling and gains an accuracy comparable to UCCSD(T).  Its error, however, grows with the number of electrons $N_e$ due to the size-consistency error as expected. The NPE also slightly increases with the number of {\it entangled} electrons as is evident from the results of isoelectronic systems, HF (NPE of 0.91 mH) and H$_2$O (3.77 mH), for the reason that will become clear shortly. Notice a similar deterioration can be observed in unrestricted methods, but for a different reason---spin-contamination. The spin-contamination error is largest halfway to dissociation, the region where static correlation can no longer be mimicked by simply breaking spin-symmetry. Consequently, the NPEs of size-consistent UCCSD(T) are also appreciably large for all the tested systems.  MR methods (CASPT2 and MRCI) are stable in this regard because they use CASSCF as the reference, and hence yield quantitatively accurate results but with a steep  increase of the  computational cost.  UCISD suffers from both size-consistency error and spin-contamination error. Comparing the results for HF and H$_2$O, we find the significance of the size-consistency error in UCISD (approximately quantified as the difference between the UCISD and UCCSD energies) is noticeably affected by not only $N_e$ but also the degree of strong correlation. It is thus considered that the accuracy of ECISD can also depend on the number of entangled electrons, as is indeed the case.

\begin{table}[]
\footnotesize

\caption{Non-parallelity errors in mHartree.}\label{tb:NPE}
\begin{tabular}{cccccccccccccccc}
\hline\hline
\scriptsize  System &	\scriptsize UCCSD(T)&\scriptsize	CASPT2&\scriptsize MRCI & \scriptsize MRCI+Q & \scriptsize ECISD&\scriptsize ECISD+Q\\\hline
HF & 3.26 & 1.28 & 1.24 & 0.32& 0.91 & 0.13 \\
H$_2$O & 8.58 & 2.60  & 0.91 &  0.16 & 3.77 & 1.29\\
N$_2$ & 15.15 & 6.91 & 1.41 & 0.57 & 15.13 & 1.51 \\
\hline\hline
\end{tabular}
\end{table}

When size-consistent correction $\Delta E_{\rm Q}$ is introduced, the error in ECISD almost disappears in spite of the size-consistency error inherent to SUHF. 
This result is intriguing, given that $\Delta E_{\rm Q}$ is a correction to $\Delta E_{\rm SD}$ but not to $E_{\rm SUHF}$; it indicates that, while $\Delta E_{\rm SD}$ itself is size-inconsistent, it yet mitigates the size-consistency error of SUHF. As a result, not only does ECISD+Q give remarkably small NPEs, but also it is almost exact in energy along the entire nuclear coordinates for all the test cases. It is also worth noting that in all the cases its accuracy at equilibrium is of the CCSD(T) level of theory, which is the ``gold-standard'' with a ${\cal O}(N^7)$ cost.  Incidentally, MRCI+Q gives smaller NPEs than does ECISD+Q as expected, but at the cost of larger computational complexity.

 \begin{figure}[t!]
\includegraphics[width=80mm]{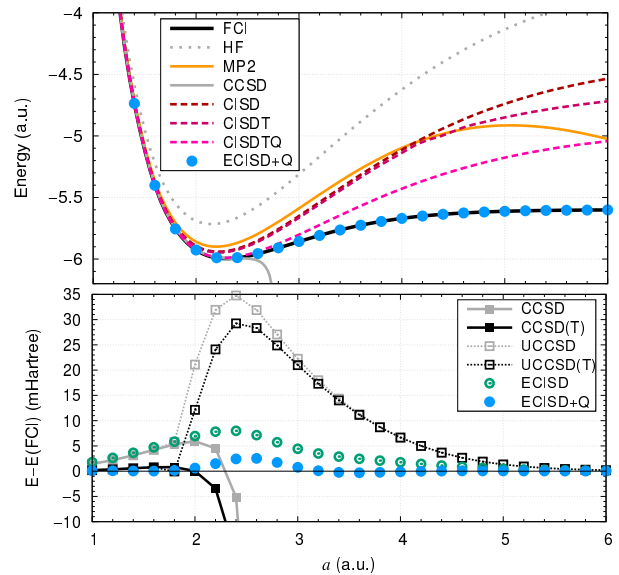}
\caption{({\it Top}) Potential energy curves of H$_{12}$ as a function of lattice parameter $a$. ({\it Bottom}) The energy errors from FCI.}\label{fig:H12}
\end{figure}
Finally, we show an even more challenging system for strong correlation: the $4\times 3$ hydrogen lattice. This system has been studied by other authors as a model system that represents a metal-insulator transition as the lattice parameter $a$ increases.\cite{Knizia13, Phillips14} When $a \rightarrow \infty$, its electronic structure exhibits very complicated spin-entanglement, and its complete description requires an active space of ($12e,12o$), which is nothing but FCI for a minimal STO-3G basis and requires 853,776 determinants. As shown in Figure \ref{fig:H12}, while many sophisticated spin-restricted and unrestricted methods break down for this system, both ECISD and ECISD+Q offer a balanced description between dynamic and static correlations. Their NPEs are found to be 8.01 and 2.84 mH, respectively. Our schemes accomplish this feat only with 1819 determinants, exactly the same number needed for single and double substitutions.

In summary, we proposed spin-extended CISD as a promising alternative to sophisticated traditional MR methods for strongly correlated systems. ECISD provides a black-box, cost-effective treatment of dynamic and static correlations in a balanced way. A remarkable accuracy was achieved when an approximate quadruples contribution is introduced through the adapted Davidson correction. 
However, ECISD+Q is an {\it a posteriori} approximation and thus works only for ground state {\it energies}, and further developments remain to be seen. We are currently working along this line.

We would like to thank Drs. Yuhki Ohtsuka, Yu-ya Ohnishi, and Motoyuki Uejima for valuable discussions and helpful comments. This work was supported in FLAGSHIP2020 by MEXT as the priority issue 5 (Development of new fundamental technologies for high-efficiency energy creation, conversion/storage and use). We are also grateful for the computer resources through the HPCI System Research project (Project ID: hp150278).

\bibliographystyle{aip}

\end{document}